\documentclass[apj]{emulateapj}

\newcommand{\simlt}{\rlap{$<$}{\lower 1.0ex\hbox{$\sim$}}}
\newcommand{\simgt}{\rlap{$>$}{\lower 1.0ex\hbox{$\sim$}}}

\newcommand{\kmsno}{{\rm km~s}\ensuremath{^{-1}}}
\newcommand{\kms}{\kmsno\ }
\newcommand{\lyano}{Ly\ensuremath{\alpha}}
\newcommand{\lya}{\lyano\ }
\newcommand{\etalno}{et~al.}
\newcommand{\etal}{\etalno\ }
\newcommand{\hseventy}{\ensuremath{h^{-1}_{70}}}
\newcommand{\Lstarno}{\ensuremath{L^\star}}
\newcommand{\Lstar}{\Lstarno\ }
\newcommand{\Zmean}{[Z]}
\newcommand{\dv}{$(\Delta v)_{\lya}$~}
\newcommand{\NHI}{N$_{\rm HI}$} 
\newcommand{\NOVI}{N$_{\rm OVI}$} 
\newcommand{\cd}{cm$^{-2}$} 
\newcommand{\FUSE}{{\it FUSE}}
\newcommand{\HST}{{\it HST}}

\begin{document}

\title{The Galaxy Environment of \ion{O}{6} Absorption Systems}
\author{John T. Stocke, Steven V. Penton, Charles W. Danforth, J. Michael Shull, \\ 
Jason Tumlinson\altaffilmark{1}, and Kevin M. McLin\altaffilmark{2}}
\affil{Center for Astrophysics and Space Astronomy, Department of Astrophysical
\& Planetary Sciences, Box 389, University of Colorado, Boulder, CO 80309}
\altaffiltext{1}{Now at Dept. of Astronomy, Yale University, New Haven, CT 06520-8101} 
  
\altaffiltext{2}{Now at Dept.\ of Physics \& Astronomy, Sonoma State 
   University, 1801 E. Cotati Ave., Rohnert Park, CA 94928}

\email{stocke, spenton, danforth, mshull@casa.colorado.edu, 
jason.tumlinson@yale.edu, mclin@universe.sonoma.edu}

\shorttitle{O~VI Absorber-Galaxy Associations}
\shortauthors{Stocke \etalno}

\begin{abstract}

We combine a \FUSE\ sample of \ion{O}{6} absorbers ($z < 0.15$) with a 
database of 1.07 million galaxy redshifts to explore the relationship
between absorbers and galaxy environments. All 37 absorbers with 
\NOVI\ $\geq 10^{13.2}$ \cd\ lie within 800~\hseventy~kpc of the nearest 
galaxy, with no compelling evidence for \ion{O}{6} absorbers in voids.  
The \ion{O}{6} absorbers often appear to be associated with environments 
of individual galaxies.  Gas with $10\pm5$\% solar metallicity 
(\ion{O}{6} and \ion{C}{3}) has a median spread in distance of 
350--500~\hseventy~kpc around \Lstar galaxies and 200--270~\hseventy~kpc 
around 0.1\Lstar galaxies (ranges reflect uncertain metallicities of 
gas undetected in \lya absorption). In order to match the \ion{O}{6} 
line frequency, $(d{\cal N}/dz) \approx 20$ for \NOVI\ $\geq 10^{13.2}$
\cd, galaxies with $L \leq 0.1$\Lstar must contribute to the cross section. 
Lyman-$\alpha$ absorbers with \NHI\ $\geq 10^{13.2}$ \cd\ cover $\sim50$\% 
of the surface area of typical galaxy filaments. Two-thirds of these show 
\ion{O}{6} and/or \ion{C}{3} absorption, corresponding to a 33--50\% 
covering factor at $0.1 Z_{\odot}$ and suggesting that metals 
are spread to a maximum distance of $800$~\hseventy~kpc, within 
typical galaxy supercluster filaments. Approximately 50\% of the 
\ion{O}{6} absorbers have associated \lya line
pairs with separations \dv $= 50-200$ \kmsno. These pairs could 
represent shocks at the speeds necessary to create copious \ion{O}{6},    
located within 100~\hseventy~kpc of the nearest galaxy and accounting 
for much of the two-point correlation function of low-$z$ \lya forest 
absorbers.

\end{abstract}

\keywords{intergalactic medium --- quasars: absorption lines ---
ultraviolet:galaxies --- galaxies: dwarfs --- galaxies: starbursts}

\section{Introduction}\label{sec:intro}

Warm, photoionized gas in the intergalactic medium (IGM) contains
virtually all the baryons in the universe at $z>2$.  With the 
growth of large-scale structure at later cosmic times, much of this 
gas cools into clumps and galaxies, while other gas is shock-heated 
to temperatures of $10^{5-7}$~K \citep{Cen99,Dave99}.  Even at 
$z \sim 0$, approximately 30\% of all baryons still reside in 
the warm ($T \approx 10^4$~K) photoionized \lya forest (Penton, 
Stocke \& Shull 2004).  Another 30-40\% of the baryons may reside 
in even hotter gas ($T = 10^{5-7}$~K), the ``warm-hot IGM'' or 
WHIM \citep{Cen99,Dave99,Nicastro05}.  

In a series of papers using moderate-resolution spectrographs
aboard the {\it Hubble Space Telescope} (\HST), the Colorado group has 
identified a sample of nearly 200 \lya absorbers 
\citep{PaperI,PaperII,PaperIV}, hereafter denoted
Papers I, II, and IV.   The \lya absorption line is sensitive 
to warm, photoionized gas, and high sensitivity \HST\ spectra with
10--20 \kms resolution can detect absorbers with \NHI\ $\geq 10^{12.5}$ \cd. 
Hotter, shock-heated gas is less easily detected, because the 
Lyman lines become weak and broad with increasing temperatures, while
higher ionization metal lines (\ion{C}{3}, \ion{C}{4}, \ion{O}{6}, 
\ion{Ne}{8}) require gas of significant metallicity 
($\geq$ 3\% solar metallicity) in order to be detectable.

The search for the WHIM has now begun, both in the soft X-rays and 
with the sensitive ultraviolet \ion{O}{6} resonance lines 
at $z \geq 0.12$ with the Space Telescope Imaging Spectrograph (STIS) 
on \HST\ \citep{Tripp00,Savage02,Tripp05} and at $z \leq 0.15$ with the 
{\it Far Ultraviolet Spectroscopic Explorer} (\FUSE) as described by 
\citet{Danforth05}, hereafter denoted DS05. At the present time, the 
\HST\ and \FUSE\ approaches have each netted $\sim 40$ \ion{O}{6} absorbers.
An analysis (DS05) of the absorber frequency per unit redshift
suggests that $\sim 5$\% of all local baryonic mass is in WHIM at 
$10^{5-6}$~K, assuming that 20\% of the oxygen is in \ion{O}{6} and 
that [O/H] $\approx -1$.  This baryon assessment assumes that all 
\ion{O}{6} absorbers are formed in collisionally ionized gas
\citep{Danforth05, Savage05}, although photoionization models can reproduce 
some of the observed line strengths, widths, and ratios \citep{Tripp01}.  
For a few \ion{O}{6} absorbers, photoionization 
of gas with very low density ($\sim 10^{-5}$ cm$^{-3}$) and large sizes 
($\sim 1$ Mpc) can provide a match to the observables 
\citep{Prochaska04,Savage02,Tripp01}. Thus, while the \ion{O}{6} systems
account for $\sim 5$\% of baryons, not all of them are necessarily at
temperatures identified as WHIM ($10^{5-7}$~K).  The hotter WHIM is 
detectable only through weak X-ray absorption lines from highly-ionized 
species such as \ion{O}{7}, \ion{O}{8}, \ion{N}{6}, and \ion{N}{7} 
\citep{Nicastro05} or possibly very broad \lya lines. 
The \ion{O}{7} X-ray detections are still too few to establish an accurate 
line density and baryon fraction, although it could be as large as 
suggested by simulations \citep{Cen99,Dave99,Dave01}.

The simulations have proven fairly accurate in predicting the baryon 
content of the warm, photoionized IGM. Some 30\% of all baryons were 
predicted by \citet{Dave99} to be in the $10^4$~K phase at $z=0$, and 
$29 \pm 4$\% of the baryons have been accounted for in the low-$z$ 
\lya forest surveys (Paper~IV).  Simulations also predict that 
30--40\% of all baryons should reside in WHIM gas at the present epoch 
\citep{Dave01}.  An important factor in verifying this prediction is 
WHIM detectability. Since highly ionized metal lines are its most sensitive 
tracers, metallicity becomes a threshold factor. \citet{Viel05}
use numerical simulations to show that when the overdensity in dark 
matter and/or galaxies exceeds $\delta \geq 10$, a mean metallicity 
above 10\% solar is expected in the IGM.  Naturally, this result depends 
on the strength of galactic winds and their ability to spread metal-enriched 
gas into their surroundings.  Mechanical and chemical feedback from 
galaxies may be more important for the WHIM than for warm, 
photoionized gas. Scenarios for producing \ion{O}{6} include outflowing 
winds from galaxies impacting IGM clouds or pristine IGM clouds falling 
into galaxies or galaxy groups \citep{Shull03,Tumlinson05}. Numerical 
simulations place WHIM gas
closer to galaxies (50--500~\hseventy~kpc) than the bulk of the warm,
photoionized gas at 300--3000~\hseventy~kpc \citep{Dave99}. 

Low-$z$ galaxy surveys along sight lines where \lya absorption lines were
detected by \HST\ have been conducted by several teams \citep[hereafter
denoted Paper~III]{Morris93,Lanzetta95,Chen98,Chen00,Chen05,Aracil02,
Prochaska04,Tripp98,Impey99,PaperIII} and have generally 
confirmed predictions for absorber-galaxy associations. 
In Paper~III we used 46 \lya 
absorbers in regions surveyed at least to \Lstar galaxy depths to find 
that the median distance from a \lya absorber to the nearest galaxy is 
500~\hseventy~kpc. Some absorbers were found within 100~\hseventy~kpc of 
galaxies, while others were Mpc away in galaxy voids \citep{McLin02}. 
Our latest sample of 138 low-redshift \lya forest absorbers in regions 
well surveyed for galaxies finds that the stronger \lya forest lines 
(\NHI\ $\geq 10^{13.2}$ \cd) are more tightly correlated with galaxies 
(Paper~IV, Stocke, Shull, \& Penton 2005) than are weaker absorbers. 
A recent paper by \citet{Chen05} confirms this result, using
data on the PKS~0405-123 sightline.  The absorbers also cluster  
more strongly with each other, as confirmed by the two-point correlation 
function (TPCF) of low-$z$ \lya absorption lines (Paper~IV). Simulations
by \citet{Dave99} predict that the WHIM can produce \lya absorption 
with \NHI\ $\geq 10^{13.2}$ \cd, with significantly smaller nearest-galaxy 
distances than for diffuse, photoionized gas.

In this paper, we investigate the relationship between galaxies and the
\ion{O}{6} absorbing clouds, at the low-temperature end of the WHIM. We 
use the DS05 sample of 40 \ion{O}{6} absorbers discovered by \FUSE\ 
\citep[see also][]{Danforthprep}, which is currently the best sample to 
investigate the relationship between the WHIM and galaxies. At $z\leq 0.15$, 
wide-field surveys for relatively bright galaxies (limiting $m_B$ of 
16.5 to 19.5) can be used to search for absorber-\Lstar galaxy pairs 
out to $z_{\rm max}=0.04-0.15$. The deepest versions of current galaxy surveys 
are well matched to the sample of DS05.

In \S~2, we briefly describe the \ion{O}{6} absorber sample and the galaxy 
redshift surveys used to define the relationship between \ion{O}{6} and 
galaxies. In \S~\ref{sec:associations} 
we describe the results of this correlation study and compare the galaxy 
environment of \ion{O}{6} absorbers to \lya absorbers in general. 
We also discuss a subset of \ion{O}{6} 
absorbers, whose associated \lya lines are paired with another \lya line at 
\dv $\leq 200$ \kmsno. 
These \ion{O}{6} absorbers in \lya line pairs are the best candidates  
for collisionally ionized WHIM. In \S~\ref{sec:implications} we discuss 
our results and place the \ion{O}{6} absorbers into the context of galaxies,
feedback/infall models and the spread of metals into the IGM.
Section~\ref{sec:conclusions} summarizes our most important conclusions.

\section{Absorber and Galaxy Samples}\label{sec:samples}

The \ion{O}{6} absorber sample is described in DS05 and in more
detail in \citet{Danforthprep}. \FUSE\ spectra were searched for \ion{O}{6}
doublet absorptions at the redshifts of strong \lya absorbers 
($W_{\lambda} \geq 80$~m\AA; \NHI\ $\geq 10^{13.2}$ \cd) found in \HST/GHRS
or STIS spectra. Most of the \lya absorbers were discussed in 
Papers I, II, and IV, augmented by STIS medium-resolution echelle 
(E140M) spectra of several bright targets observed by other investigators. 
Echelle sight lines were included when high-quality \FUSE\ data are available 
for the same targets. DS05 used \FUSE\ spectra of 31 AGN targets to search 
for \ion{O}{6} absorption associated with 129 known \lya absorbers with
$W_{\lambda} \geq 80$ m\AA.  Of their 40 detections of \ion{O}{6} at 
$\geq 4\sigma$ level, we use only the 37 detections with \NOVI\ $\geq 10^{13.2}$ 
\cd\ for our \ion{O}{6}-detected sample, to ensure 90\% completeness.  
Our detections in \NOVI/\NHI\ correspond to (O/H) abundances of 
$\sim9$\% solar (DS05); see also Prochaska \etal\ (2004) for a similar 
metallicity estimate.  The \ion{O}{6} non-detection sample includes only 
those 45 absorbers \citep{Danforthprep} with 4$\sigma$ upper limits at 
\NOVI\ $\leq 10^{13.2}$ \cd, so that 
the \ion{O}{6} properties of these two subsamples are disjoint. 
If the 82 well-observed absorbers in the DS05 sample are indicative of the
parent population, approximately 45\% (40/82) of all higher column density 
\lya absorbers have detectable \ion{O}{6} absorption. 
\citet{Danforthprep} estimate that 55\% of the \lya absorbers at 
\NHI\ $\geq 10^{13.2}$ \cd\ have associated \ion{C}{3} $\lambda977$ 
absorption.  

Bright-galaxy redshift surveys available from several groups now include 
positions and redshifts for over $10^6$ galaxies. To assemble the final
galaxy catalog, we began with 494,000 galaxies in 
the 7~Jan~2005 version of the Center for Astrophysics Redshift Survey 
\citep[ZCAT;][]{Huchra90,Huchra95,Huchra99}.  To this catalog we added 
data from other surveys, bringing the total to 1.07 million galaxies.
We made the following additions and changes:  
\begin{enumerate}
\item{Replaced the ZCAT Veron-Cetty \& Veron catalog version~9 entries with the
current catalog entries \citep[Version 11,][]{Veron11}.}

\item{Replaced the ZCAT Sloan Digital Sky Survey (SDSS) early
data release (EDR) and data release 1 and 2 (DR1 and DR2) entries
of $\sim 138,000$ galaxies \citep{SDSS-EDR,SDSS-DR1,SDSS-DR2} with 
the SDSS-DR4 catalog ($\sim 550,000$ galaxies; \url{http://www.sdss.org/dr4}).}

\item{Replaced the ZCAT Two-degree Field Galaxy
Redshift Survey (2dFGRS)-EDR entries \citep[$\sim 100,000$ galaxies;][]{2dFGRS-EDR}
with the 2dFGRS final data release \citep[$\sim 250,000$ galaxies;][]{2dFGRS-FDR}.}

\item{Replaced the $\sim 17,000$ ZCAT Six-degree Field Galaxy Redshift
Survey (6dFGRS)-EDR entries with $\sim 65,000$ entries from 6dFGRS-DR2 
with well-determined redshifts \citep{6dFGRS-DR2}.}
\end{enumerate}

These surveys were scrutinized for duplications, and several thousand duplicates
were removed before we cross-correlated absorber-galaxy locations. In addition
to the wide-field surveys named above, our group has conducted 
pencil-beam galaxy surveys along some of these specific sight lines
\citep{McLinthesis,McLin02}, adding several hundred galaxy redshifts to our 
final catalog. To determine the depth along each sight line to
which galaxy redshifts are available down to a given galaxy luminosity
\citep{Marzke98}, we relied on the published apparent B-magnitude 
limits for each survey and a $V/V_{\rm max}$ test, since more than one survey 
contributes along most sight lines. We applied K-corrections from \citet{Oke68}, 
appropriate for giant elliptical galaxies.  In all cases, these corrections
reduced the distance coverage based upon the $V/V_{\rm max}$ tests. 
The published 2DF and 6DF limiting magnitudes are B = 19.5 and 16.8 respectively.
The SDSS claims completeness to $r = 17.8$ for the spectroscopic survey. 
However, not all galaxies measured photometrically above that limit were
targeted with fibers in any one field, because fibers cannot be placed closer 
than $\sim 55\arcsec$ apart. \citet{Strauss02} determine that the overall 
SDSS is over 90\% complete for galaxies brighter than $r = 17.8$, with 
completeness reaching 99\% for fields without bright or closely-spaced 
galaxies, which represent about 10\% of the main sample. A similar circumstance 
could be important in some 2DF fields.  Several other surveys, including 
McLin (2002), have various magnitude limits and completeness levels depending on 
the specific sight line observed. 

We obtain good agreement for regions with completeness to \Lstar or better,
based on the quoted magnitude limit and the $V/V_{\rm max}$ test.
Absorbers were rarely found close to the \Lstar completeness depth,
although the $cz$=36,021\kms absorber toward 3C~273
lies right at the \Lstar completeness limit, based upon a $B = 19.0$
limit set by \citet{Morris93} and on our $V/V_{\rm max}$ test. 
Information about the galaxy environment around this absorber is uncertain, 
so to be conservative we have eliminated this single absorber from our 
sample.  This leaves 23 \ion{O}{6} absorbers and 32 \ion{O}{6} non-detections 
that lie in regions surveyed at least to \Lstar depths. For comparison, we 
use the stronger half of our entire \lya absorber sample, composed of 69 absorbers 
with {\it W}$_{\lambda} \geq 68$ m\AA\ (\NHI\ $\geq 10^{13.2}$ \cd) in regions
surveyed at least to \Lstar depths. A smaller number of absorbers in these
samples have been surveyed to 0.1\Lstar depths: \ion{O}{6} detections (9), 
\ion{O}{6} upper-limits (8), \lya stronger half-sample (20), and \lya weaker 
half-sample (19).

In order to determine the 3-dimensional distance between absorbers and galaxies,
we assum a ``retarded Hubble-flow'' model \citep{Morris93} as we have done in
our previous analyses (Paper~III, Paper~IV). In the current analysis, we have 
increased the peculiar velocity difference from $\pm300$\kms to $\pm500$\kms 
for which we consider the galaxy and \ion{O}{6} absorber to be 
at the same redshift distance.  We made this adjustment because we have
found that the \ion{O}{6} absorption is sometimes not at the same redshift 
as the associated \lya \citep{Shull03, Tumlinson05}. This peculiar velocity 
allowance, while guided by the velocity dispersion in groups of galaxies 
and galaxy filaments, is somewhat arbitrary \citep[see][for justifications 
of allowances of $\pm 250-500$ \kmsno]{Lanzetta95,Impey99}, but it does not
affect our final results.

\section{\ion{O}{6} Absorber-Bright Galaxy Associations}
\label{sec:associations}

In order to maximize the information available to study \ion{O}{6}
absorber-galaxy associations, we adopt two approaches to the merged-galaxy 
redshift database.  In our first approach, we construct 
luminosity-limited galaxy catalogs, complete along AGN sight lines to 
either \Lstar or 0.1\Lstarno.  We specifically exclude all fainter galaxies, 
even when their redshifts are known.   
Using the methods described in \S~2, we determine the maximum distance 
along each sight line to which galaxy redshift data are complete; 
regions complete to 0.1\Lstar are a small subset of the \Lstarno-complete 
regions.  Because few absorbers lie in regions surveyed 
below 0.1\Lstarno, this is the lowest galaxy luminosity limit that permits 
a sufficient sample size to draw statistical inferences.  

Our second approach uses all galaxies observed along each sight line for 
absorbers at distances closer than the \Lstar or 0.1\Lstar limits. 
Although galaxy catalogs are well defined with a single luminosity limit, 
they do not use all the available redshift information. Typically, the 
redshift survey data are limited in apparent magnitude along any one sight 
line, so that absorbers at lower redshifts are surveyed to lower galaxy 
luminosities. Such an approach, using an apparent-magnitude limited sample, 
is similar to that employed by \cite{Chen05} to construct a galaxy-absorber 
cross-correlation function. 
The shortcoming of the second approach is that each absorber has a different 
luminosity completeness limit.  In neither approach do we use absorbers in 
regions surveyed incompletely for galaxies. We term these two approaches as: 
(1) ``complete to \Lstar or 0.1\Lstarno"; or (2) ``all galaxies".

\begin{figure*}[htb]
\epsscale{0.67}
\plotone{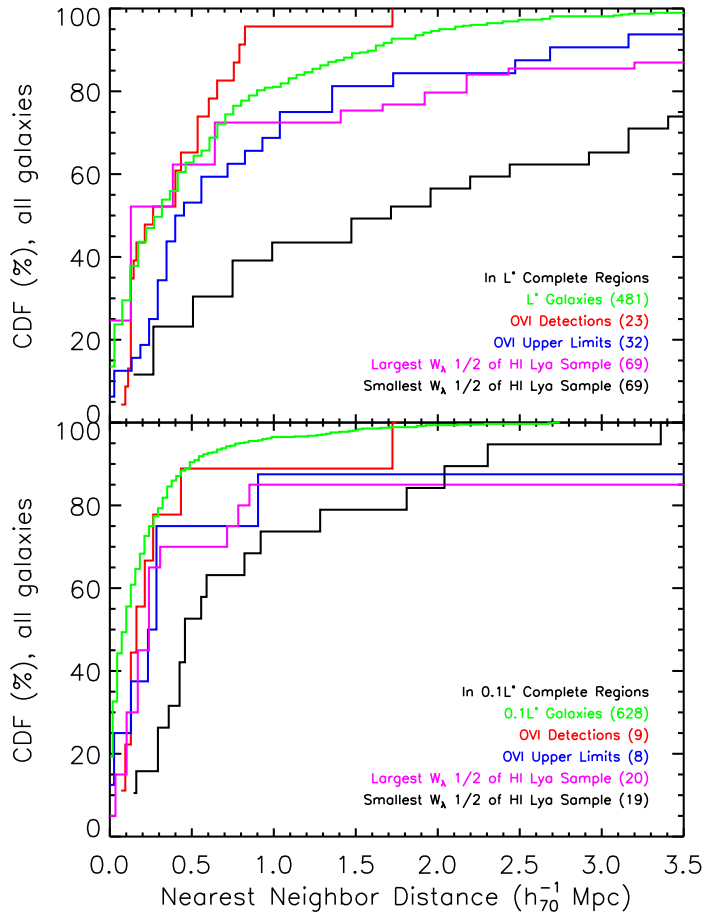}
\caption{\label{CDF} Cumulative distribution functions (CDFs), using the
``all galaxies" approach of \S~3) showing the
distance between various types of absorbers and the nearest known galaxy in
regions around absorbers, complete at least to \Lstar (top) and 0.1\Lstar
(bottom). Also shown in green are the CDFs for the distance between galaxies in
these regions.}
\end{figure*}

Figure~\ref{CDF} displays the cumulative distribution function (CDF) of 
the nearest-neighbor distances for several different samples using the 
all-galaxies approach. The CDFs in the top panel of Figure~\ref{CDF} show 
those absorbers in regions surveyed at least down to \Lstar luminosities; 
the bottom panel shows the 0.1\Lstarno-complete regions.  These CDFs are 
shown as the percentage of galaxies vs. distance between the 
nearest galaxy and the following objects:  
(1) each galaxy at or above the limiting luminosity (green line); 
(2) each \ion{O}{6} absorber (red line); 
(3) \lya absorbers with a sensitive \ion{O}{6} non-detection (blue line); 
(4) stronger \lya absorbers from Paper ~IV ({\it W$_{\lambda}$} $\geq 68$ m\AA;  
     magenta line); 
(5) weaker \lya absorbers from Paper~IV ({\it W$_{\lambda}$} $< 68$ m\AA;
    black line). 
The \ion{O}{6} sample 
of DS05 was drawn exclusively from absorbers in the \NHI\ range of the stronger
half-sample [W$_{\lambda}$(\lyano) $\geq 80$ m\AA].  The sample size in each CDF 
is shown in parentheses. Despite the fact that this \HST-derived sample of
absorbers is at very low redshift, the number of absorbers in regions
surveyed at least to 0.1\Lstar (M$_B\approx-17$) is still quite modest.

Table~\ref{nntable} compares our two approaches, using the median
nearest-neighbor distance for samples displayed in Figure~\ref{CDF}. 
If the sample size is an even number of absorbers, we report the mean of
the two median numbers. Because the nearest-neighbor distributions are
skewed, the mean is considerably larger than the median in all cases.
The mean distance between \Lstar galaxies in our absorber regions is 
2.15~\hseventy~Mpc, while the median is 1.78~\hseventy~Mpc. The sample 
sizes in Table~\ref{nntable} are shown in columns 4 and 7.

\subsection{Galaxy Statistics of \lya Absorbers}
\label{sec:lyaabsorbers}

The current \lya absorber sample is larger than that used previously to
analyze absorber environments. The CDFs in Figure~\ref{CDF} and the median 
values in Table~\ref{nntable} are based on 138 absorbers, compared to 46 absorbers 
in \Lstarno-surveyed regions in Paper III, but our new results are consistent with 
those found previously.  Nevertheless, it is worthwhile to revisit 
the basic results found in Paper III for \lya absorbers before investigating the 
\ion{O}{6} subset. 
The \lya absorbers in the stronger half-sample are more tightly correlated 
with galaxies than the weaker half-sample.  In Table~\ref{nntable}, the median 
distance to the nearest 0.1\Lstar galaxy is over three times larger for the 
weaker absorbers.  Most of these absorbers lie within a few hundred kpc of a 
bright galaxy, a location consistent with a filament of galaxies.  Yet the 
stronger half-sample also includes 10 \lya absorbers ($\sim 15$\%) in galaxy 
voids, with nearest bright-galaxy neighbors $\geq 3$~\hseventy~Mpc away
(see Paper III and Drozdovsky et al.\ 2005) for a justification of this limiting 
distance\footnote{In this paper we use the terms ``filament'' and ``void'' 
generically to define regions where galaxies exist and those where they do not
(Vogeley, Geller \& Huchra 1991; Slezak, de Lapparent \& Bijaoui 1993), 
regardless of the details of the local galaxy density and its 
extent and shape in space. Thus, we use the term filament to refer to both 
spherically-shaped galaxy groups and to long strings of gravitationally 
unbound galaxies.  While distances to the nearest galaxies in filament locations
depend strongly on the limiting luminosities of the galaxies surveyed, this
is much less true of void regions, as long as galaxies are surveyed at least
down to \Lstar luminosities \citep{Park05,Hoyle05,Szomoru94}.}.

The weaker half-sample has a CDF and median nearest-galaxy distances
similar to those found in numerical simulations of photoionized intergalactic
gas \citep{Dave99}. To best compare with the simulations, we use the median
values in 0.1\Lstar regions for all galaxies (column~6 in
Table~\ref{nntable}).  Dav\'e (private communication) estimates that the 
\citet{Dave99} simulations identify galaxy locations down to 
0.025 \Lstarno. The median nearest-neighbor distance for simulated
photoionized absorbers is 600~\hseventy~kpc, compared to 480~\hseventy~kpc
observed for the weaker half sample.  This is a reasonable match,
considering that the limiting galaxy luminosities of the observations and 
the simulations are not precisely matched and that some photoionized absorbers 
may be present in the stronger half-sample. 

In the entire \lya sample, $25\pm 4$\% of \lya absorbers lie in voids, 
more than 3~\hseventy~Mpc from the nearest galaxy and consistent with the 
$22\pm8$\% found in our smaller sample (Paper~III). 
This fraction is also consistent with the $\sim 20$\% of photoionized absorbers 
found in voids by the simulations. However, in both simulations and observations, 
the pathlengths may not be the same through voids and superclusters. 
In our current sample, the relative pathlengths (43\% through voids, 57\% 
through superclusters) are probably representative of the local universe,
since only four of our 31 targets (Mrk~335, Mrk~421, Mrk~501, I~Zw~1) 
were chosen to lie behind well-known galaxy voids.
The percentage of void absorbers drops to $\sim$10\% for the stronger absorbers 
investigated for the presence of O~VI absorption in this paper; this is shown in
Figure 1 for both \Lstar and 0.1 \Lstar samples.  

The median distances between \lya absorbers and \Lstar galaxies (Table 1) are 
so large (1.1--2.2 \hseventy~Mpc) that is implausible to associate absorbers 
with individual \Lstar galaxy halos. This is because the median distance 
(345~\hseventy~kpc) between an \Lstar galaxy and its nearest galaxy neighbor 
is a small fraction of these values, and we usually find 
lower luminosity galaxies between the absorber and the nearest \Lstar galaxy. 
Furthermore, the median distance between any two $\geq 0.1\Lstarno$ galaxies 
(245~\hseventy kpc) is smaller than that between an absorber and its nearest 
$\geq$\Lstar or $\geq 0.1$\Lstar galaxy neighbor (625 and 335 \hseventy~kpc,
respectively).  This also makes it difficult to assign a specific galaxy at 
$\geq0.1$\Lstar to an individual absorber.
Comparing the CDFs for the \Lstarno-complete and 0.1\Lstarno-complete samples, 
we find that the median nearest-neighbor distances for all absorber samples 
scale by the expected amount if the galaxies are located randomly with respect 
to absorbers.  For example, by extending the galaxy surveys
fainter, the galaxy density is increased by an amount given by standard
luminosity functions \citep{Marzke98,Blanton03}, and the nearest-neighbor 
distance is decreased by approximately the cube root of that amount.

These last two statistical results are the primary evidence that low column 
density \lya absorbers are not easily ascribed to individual galaxies but 
rather to supercluster filaments \citep{PaperIII,Stocke05}.  Gas 
in these absorbers is either of primordial origin or it comes from the 
cumulative debris of many galaxies in the filament. 
\citet{Impey99} came to similar conclusions, using a smaller sample extending 
to galaxies with $M_B \leq -16$ but confined to the Virgo Cluster. 
Although their conclusion may be biased by the special conditions 
in the Virgo Cluster (see \S~\ref{sec:implications}), only three of our
31 sight lines pass through any portion of the Virgo Cluster.
Assigning \lya absorbers to supercluster filaments is also consistent with
numerical simulations of the low-$z$ \lya forest \citep{Dave99}.  
Owing to the incompleteness of the galaxy survey data at $L < 0.1$\Lstarno, 
we cannot rule out the possibility that some \lya absorbers are associated 
with a nearby galaxy fainter than 0.1\Lstar (see column~6 in Table~\ref{nntable}, 
Table~\ref{assoctable}, and Stocke et al. 2005).  

\subsection{Galaxy Statistics of \ion{O}{6} Absorbers}\label{sec:OVIabsorbers}

Using the ``all-galaxies" approach in Table~\ref{nntable} and Figure~\ref{CDF},
we find that the median distances from an \ion{O}{6} absorber to the nearest
galaxy in any region complete to \Lstar or 0.1\Lstar are comparable to 
the distance between galaxies in these regions. The \ion{O}{6} absorber 
CDFs in Figure~\ref{CDF} are the ones that either overlap the galaxy CDF or 
lie to the left of it.  In many cases, it may be possible to associate 
individual \ion{O}{6} absorbers with individual galaxies, although
many of these galaxies are likely to be fainter than
0.1\Lstar (see absolute magnitudes of nearest galaxies in Table~\ref{assoctable}).
The nearest-neighbor galaxy distances for \ion{O}{6} absorbers are about
twice as large as predicted for shock-heated gas by numerical simulations.
For instance, the median observed distance to 0.1\Lstar galaxies is 180
\hseventy~kpc (Table~\ref{nntable}, column~6) compared to 100~\hseventy~kpc
in the simulations \citep{Dave99}.  A subset of \ion{O}{6} absorbers
in \lya pairs has even smaller nearest-neighbor distances, which are better
matched to the galaxy distances in simulations. These absorbers will be defined
and discussed in the next section.

Because the \ion{O}{6} detections and non-detections were drawn from the 
stronger half of the \lya absorber sample, it is not surprising that the CDFs 
in the top panel of Figure~\ref{CDF} for these three samples are similar. 
The \ion{O}{6} detections are found almost exclusively at distances $< 800$
\hseventy~kpc from galaxies in \Lstarno-complete regions and $< 400$
\hseventy~kpc from galaxies in 0.1\Lstar regions.  The one exception to
this statement is a problematical case, discussed below. The median distances 
between \ion{O}{6} absorbers and the nearest $\geq$\Lstar and $\geq0.1$\Lstar 
galaxy are 625 \hseventy~kpc and 335~\hseventy~kpc respectively (Table~\ref{nntable}, 
columns 2 and 5). The medians provide plausible estimates for the typical 
distance that metals have been spread away from bright galaxies, 
even if the nearest bright galaxy is not the source of the metals.
We refine these distances using covering-factor estimates in 
\S~\ref{sec:implications}.

One-third of all \ion{O}{6} non-detections lie more than 1~\hseventy~Mpc 
(800~\hseventy~kpc) from 
the nearest \Lstar (0.1 \Lstarno) galaxy, a greater distance than any \ion{O}{6}
detection. This result is expected because of the finite distances that metals 
can be spread away from galaxies in the current-day universe \citep{Tumfang05}. 
Gas blown 1~\hseventy~Mpc from its source 
would require a mean wind speed of 500\kms acting for 2 Gyr. Outflowing winds 
from galaxies in the present-day universe are expected to have mean speeds 
substantially less that this amount \citep{Martin03, Stocke05}, and 2 Gyr
is much longer than the typical ($10^8$-yr) duration for wind activity.  
More distant absorbers could contain metals, but be too cool 
to produce \ion{O}{6} absorption.  Such an effect is expected, since warm 
photoionized absorbers in the simulations tend to exist farther from galaxies 
than the collisionally-ionized WHIM \citep{Dave99}. We will propose a solution 
to this uncertainty in \S~\ref{sec:implications}.

The CDF for the \ion{O}{6} absorbers includes only one absorber in the DS05
\ion{O}{6} sample with no nearby ($<1$~\hseventy~Mpc) galaxy. 
The most isolated \ion{O}{6} absorber is the 13,068 \kms absorber 
toward Ton~S180, located 1.7~\hseventy~Mpc from the nearest 
bright galaxy.  This system is part of a complex of two \lya absorption pairs 
(Paper IV) at velocities $cz$ = (12,192, 13,068 \kmsno) and (13,515, 13,681 
\kmsno). A figure showing these \lya pairs is given in \citet{Shull02}. 
We detect \ion{O}{6} in just one component of each pair, at 13,068 \kms
and 13,681 \kmsno. The higher-redshift absorber pair has a nearest galaxy 
285~\hseventy~kpc away at $cz = 13,798$ \kmsno, a typical nearest-neighbor 
distance for \ion{O}{6} detections.  However, in our ``retarded Hubble flow'' 
model, this galaxy and another located 3.3~\hseventy~Mpc from the 
$cz$=13,068\kms absorber, could be at comparable physical distances.   
Given the potential physical relationship between this complex of four
\lya absorptions, we can not exclude the possibility that the 13,068\kms
absorber has a large peculiar velocity and is actually only 
$\sim$300~\hseventy~kpc from the same bright galaxy identified with the 
$cz \approx 13,000$ \kms absorbers.

We have found no compelling evidence for detectable \ion{O}{6} absorbers in
voids.  However, there is significant evidence that \ion{O}{6} absorbers
lie within 70--400~\hseventy\ kpc of a relatively bright 
($M_B \leq -17$) galaxy. Table~\ref{assoctable} lists ten 
\ion{O}{6} absorber-galaxy pairs in the DS05 sample with total inferred 
distances $\leq 200~$\hseventy~kpc.
A large fraction of these close pairs involve low-luminosity galaxies, 
despite the fact that our galaxy survey is incomplete at L$<0.1$\Lstarno. 
This suggests that deeper galaxy surveys in regions near other 
\ion{O}{6} absorbers will find low-luminosity galaxies close to these absorbers.

\subsection{Statistics of Non-Detected \lya Absorbers}
\label{sec:nondetects}

To understand how far metals are spread from galaxies, it is insufficient to
measure only the distance between absorbers and their nearest-neighbor galaxy. 
We also do the reverse experiment, searching for \lya absorbers at the
same redshifts as galaxies located within 800 \hseventy~kpc of the sightline.   
For this census we use only those sight lines from Papers~I and IV, for
which the sensitivity function is well-determined.
Excepting wavelengths obscured by Galactic absorption lines, the sensitivity
function for \lya absorption is relatively flat as a function of wavelength
and column density. All \HST\ spectra are of sufficient signal-to-noise ratio 
to detect \lya absorption down to \NHI\ $=10^{13.2}$ \cd.  Some of the best 
spectra are able to detect \lya absorption down to \NHI\ $\approx10^{12.5}$ 
\cd\ (Paper~IV).  By searching within 800~\hseventy~kpc of all sight lines 
out to the maximum recession velocity for which \Lstar galaxies have been 
completely surveyed, we identified 16 galaxies with no accompanying
\lya absorption and 26 additional galaxies detected at \NHI\ $\leq
10^{13.2}$ \cd. In these same regions, we find 53 galaxies within $\pm
500$\kms of a \lya absorber. 

Our previous accounting in \S~\ref{sec:lyaabsorbers} assigned the \lya 
absorber only to the nearest galaxy. In the current accounting, we should
use only the nearest galaxy within 500\kms as the ``non-detected'' galaxy. This
method is equivalent to assigning the detections and non-detections to a galaxy
supercluster filament, not to an individual galaxy, an interpretation supported
by the data presented above. Galaxy filaments have typical thicknesses 
of 5--7~\hseventy~Mpc \citep[estimate for the ``Great Wall'' by][]{Ramella92},
so in a pure Hubble flow galaxies in a filament can be separated in
recession velocity by 350--500 \kmsno.  We thus identify  
50 galaxy filaments (individual galaxies or groups of galaxies
within 500 \kmsno) located $\leq 800$~\hseventy~kpc from sight lines
in \Lstarno-complete regions. These filaments typically contain 1-6 galaxies 
within 800~\hseventy~kpc of the sight line. Of these fifty galaxy
filaments (\lya detections and non-detections), half lie within $\pm
300$\kms of a \lya absorber at \NHI\ $\geq 10^{13.2}$ \cd. Of the
remaining half, 15 are detected at lower \NHI, two are possibly detected
at 3--4 $\sigma$ significance, and eight are not detected. Thus, we infer 
a 50\% covering factor for gas at \NHI\ $\geq 10^{13.2}$ \cd\ in galaxy filaments. 
Down to our \lya detection limits, at least 80\% of these filaments are
covered, which would suggest a covering factor close to unity if more 
sensitive UV spectra were available. Thirteen of the 40 filaments definitely 
detected in \ion{H}{1} have multiple \lya absorptions, although we have 
counted these \lya complexes as single absorbers in this census.
That is, the maximum covering factor of any one filament is one. 
The CDFs for galaxies with non-detected \lya absorbers are similar to
those of the stromger \lya absorbers, with median distances of 
350--500~\hseventy~kpc. Since we have restricted our investigations 
to galaxies $\leq$800~\hseventy~kpc from each sight line, both of these CDFs 
are similar to the CDF for \ion{O}{6} absorbers shown in Figure~\ref{CDF}. 

In Table~3 we list those galaxies located $\leq 200$~\hseventy~kpc 
from a sight line for which \lya absorption is not detected, as well as 
weak absorbers detected at \NHI\ $<10^{13.2}$ \cd. Three of these 
galaxies lie near the PG~1211+123 sight line \citep{Tumlinson05} 
with recession velocities placing them in one of the Virgo Cluster 
``galaxy clouds''. Other Virgo galaxies, as well as a strong \lya absorber, 
are seen at slightly higher velocities. The non-detection of 
the environment surrounding these galaxies may arise from special circumstances 
in Virgo, the only large cluster probed by this survey. The other galaxy 
with undetected \lya absorbers lies in the 3C~273 sight line, 
just beyond the Virgo recession velocities. These are small numbers, but it 
is intriguing that 3 of the 4 galaxies within 200~\hseventy~kpc of target 
sight lines showing no associated \lya absorption lie in the Virgo Cluster, 
even though only 3 of our 31 sight lines pass through this cluster.
Special conditions in Virgo may also explain the 
\citet{Impey99} result that even very low luminosity galaxies 
($M_B \leq -16$) are not obviously related to \lya absorbers.

\subsection{\ion{O}{6} in \lya Line Pairs}\label{sec:pairs}

\begin{figure*}[htb]
\epsscale{0.5}
\includegraphics[angle=90,scale=0.667]{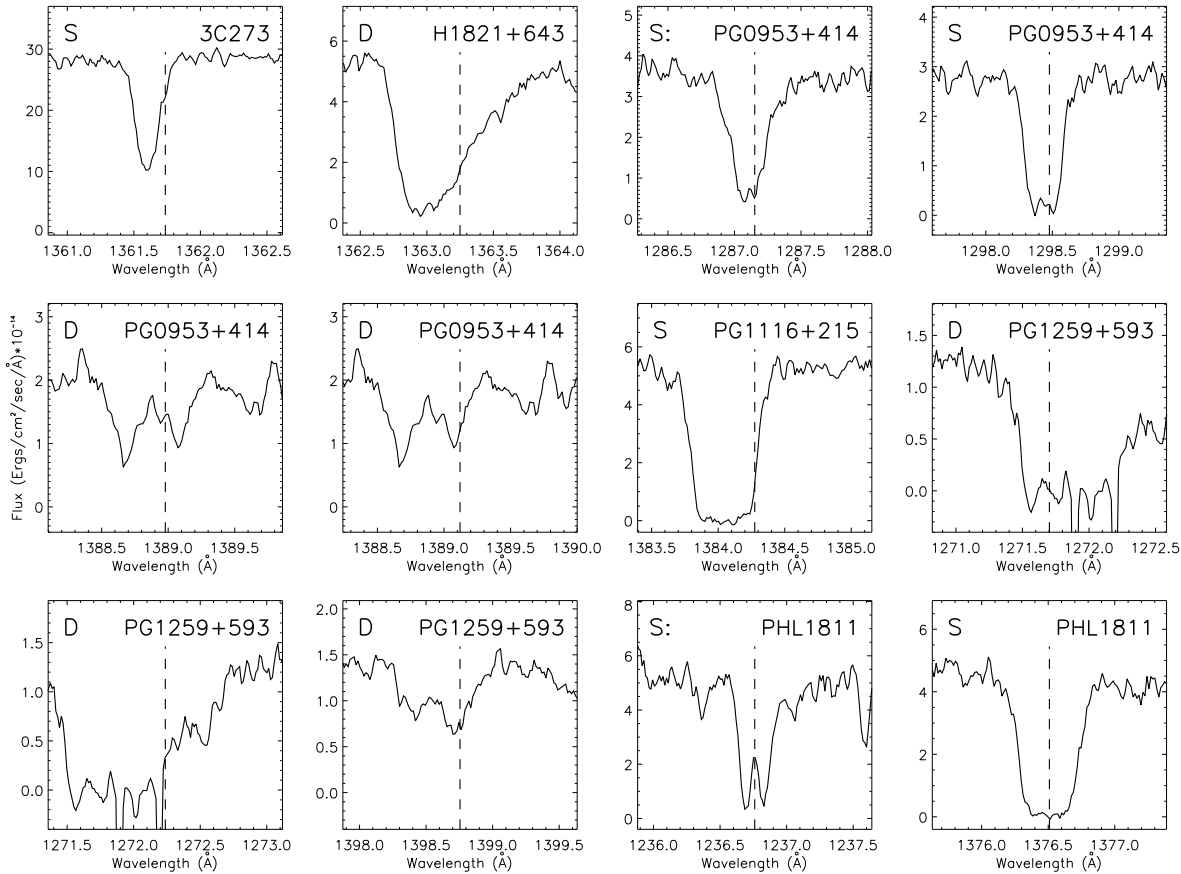}
\caption{\label{OVIlineprofiles} Examples of \lya line profiles associated 
with \ion{O}{6} absorbers in the Danforth \& Shull (2005) sample. The 
locations of the \ion{O}{6} absorption line ($\pm 0.2$~\AA\ due to errors 
in line centroiding and \FUSE\ wavelength calibration) are shown as vertical 
dashed lines. The letter in the upper left of each box is our proposed 
classification: S = single line; S: = uncertain, counted as single line; 
D = double line. We have classified the PHL~1811 \lya line as S: despite 
the apparent peak in the center of the line, because such a division
would make each \lya line implausibly narrow ($\leq 15$ \kmsno).}
\end{figure*}

As we pointed out in Paper~IV, the excess power in the two-point correlation
function (TPCF) of \lya absorption lines occurs exclusively within the 
stronger half-sample (W$_{\lambda} \geq$ 68 m\AA).  This sample contains the 
\ion{O}{6} absorber population discussed here. Thus, we expect that some of these 
\ion{O}{6} absorbers are associated with \lya absorber pairs at small velocity 
separations, as seen toward PKS~2155-304 \citep{Shull03}  
and PG~1211+143 \citep{Tumlinson05}.  The GHRS
and STIS spectra allow a reliable search for line pairs at \dv $\geq 50$\kms
for first-order spectra (Papers~II and IV) and at \dv $\geq 30$\kms for 
STIS echelle spectra. Absorbers that are blended at \lya were scrutinized at
Ly$\beta$ or higher Lyman lines using \FUSE\ spectra. Examples of \lya line
profiles from the \ion{O}{6} absorber sample are shown in
Figure~\ref{OVIlineprofiles}, including single isolated lines, clear pairs of
lines, complexes of more than two lines, and broad, asymmetric lines verified 
as pairs using higher Lyman lines. For this investigation, we group the line 
complexes with the line-pairs, but we count each complex as only a single 
line pair.

The statistical results are striking: 16 of the 37 \lya lines in the 
\ion{O}{6} sample are \lya line pairs with 
\dv $\leq 200$ \kmsno.
Only some of these \lya pairs show \ion{O}{6} absorption in both lines. 
Using all 21 other absorbers at \dv $\geq 400$\kms to determine the number 
of close line pairs arising by chance, we expect fewer than 2 pairs in any 
\dv $= 200$\kms interval. The presence of 16 \lya absorbers in close pairs 
is highly significant according to the binomial probability distribution 
(P$_B< 10^{-10}$).

As described and listed in Papers II and IV, the TPCF excess, which is seen
exclusively in the stronger half of the local \lya sample (94 members) arises 
from 10 line pairs.  We found that  11\% of the stronger half-sample has 
\dv $\leq 150$ \kmsno. Within the DS05 \ion{O}{6} sample, we found 9 \lya line 
pairs (7\% of the full DS05 sample) with \dv $\leq 150$\kms and four 
others with \dv $=150-200$ \kmsno. 
As above, the number of \lya lines (26) in pairs in the DS05 sample is 
different from the number of \ion{O}{6} absorbers (16) in pairs,
because most of the \lya pairs do not show \ion{O}{6} in both lines. 
The two samples from Paper~IV and DS05 overlap considerably, 
but they are not identical. Within the statistics, the
\ion{O}{6} in \lya line pairs can account for most of the TPCF amplitude
shown in Paper~IV. 
Using all the \lya line-pairs and complexes found in Papers II and IV, we 
have searched for associated O~VI absorption based upon our previously described 
limits (see \S~2). In this sample we find that half are detections and half 
are non-detections. Thus, we suggest that \ion{O}{6} absorbers in \lya line pairs 
may be the majority population that exhibits clustering in the low-$z$ \lya forest.

Ten of the 16 \ion{O}{6} absorbers in \lya line pairs lie in areas surveyed
for galaxies to at least \Lstar depth. Six of these have nearest-galaxy
distances $\leq 160$~\hseventy~kpc, three have nearest-galaxy distances of
200--450~\hseventy~kpc, and the last is the anomalous absorber toward
Ton~S180 discussed above. Despite the ambiguous circumstance of
the Ton~S180 absorber, the \ion{O}{6} in \lya line pairs sample
is closely associated with galaxies. With a median absorber-galaxy distance of
190~\hseventy~kpc in \Lstarno-complete regions, these absorbers lie at 
significantly less
than the median distance between galaxies in these same regions. Just as for
the \ion{O}{6} absorbers as a whole, the \ion{O}{6} absorbers with \lya line
pairs might plausibly be associated with an individual, nearby, usually faint
galaxy. Unfortunately, only three \ion{O}{6} absorbers in \lya line pairs are
in areas surveyed for galaxies at least to 0.1\Lstar depth (see
Table~\ref{nntable}), so that an accurate assessment of the nearest faint
galaxy distance is not yet possible. However, we can estimate the expected
distances for 0.1 \Lstar galaxies from these absorbers using the relative
numbers of \Lstar and 0.1\Lstar galaxies present in standard luminosity
functions \citep[e.g.,][]{Marzke98,Blanton03}. Scaling the \Lstar region
results in Table~\ref{nntable} to 0.1\Lstarno, we expect to find fainter
galaxies at a median distance of $\leq 100$~\hseventy~kpc from these absorbers.
Thus, while \ion{O}{6} absorbers in general have median nearest-galaxy
distances somewhat larger than predicted for WHIM gas by \citet{Dave99}, the
\ion{O}{6} absorbers in \lya line pairs match the 100~\hseventy~kpc median
distance to the nearest \simgt 0.025\Lstar galaxy predicted by the simulations
\citep{Dave99} for this IGM component.

In contrast to the \ion{O}{6} absorbers in \lya line pairs, \ion{O}{6}
absorbers associated with isolated single \lya lines [\dv $\geq 400$ \kmsno]  
have nearest-neighbor distances spanning the full range shown in
Figure~\ref{CDF} (70--800~\hseventy~kpc). Considering that some of these
single lines could be unidentified pairs with velocity spacings
$\leq 50$ \kmsno, we estimate that $\sim$45\% of all \ion{O}{6}
absorbers could be in closely-spaced \lya line pairs. Since \ion{O}{6}
absorbers are detected in $\sim 45$\% of all higher column density \lya forest
absorbers, this suggests that $\sim 20$\% of \lya absorbers with 
\NHI\ $= 10^{13.2-16.5}$ \cd\ are members of close line pairs, at least one of 
which has associated \ion{O}{6} absorption. This population of absorbers alone
accounts for $(d{\cal N}/dz)_{\rm OVI} \approx 10$ (5 line pairs) per unit 
redshift at $z \leq 0.15$. 

We have also searched for correlations of the nearest galaxy neighbor 
distance and the \lya pair separation, with the ``multi-phase ratio",
\NHI/\NOVI, described in DS05. This ratio can be used to 
estimate the range in contributions from photoionized (\ion{H}{1}) and
collisionally ionized gas (\ion{O}{6}) in absorbers that appear associated
kinematically.  We find no correlations between the multi-phase ratio and 
these two quantities.

The distance to the nearest-neighbor galaxy is not always a reliable
measure of the local environment.  Therefore, we have also searched our 
collective galaxy catalog for all galaxies with 
$|cz_{\rm abs}-cz_{\rm gal}|< 1000$\kms within several Mpc of each 
\ion{O}{6} absorber.  This allows us to identify nearby groups of galaxies 
and filamentary structures close to the absorber. The $cz\approx$ 17,000\kms 
group of four bright spiral galaxies surrounding an \ion{O}{6} absorber pair 
and associated complex of \lya lines toward PKS~2155-304 is an obvious example 
\citep{Shull03}. There are other examples, including the
PG~1211+143 absorbers and galaxy group at $cz \approx$ 15,300\kms 
\citep{Tumlinson05} and the $cz \approx 43,000$ \kms \ion{O}{6} absorber pair in
the PG~0953+415 sight line \citep{Savage02}. 
Statistically, we find mixed evidence for
\ion{O}{6} absorbers arising in groups of galaxies. Only $\sim 60$\% (14
of 23) of the \ion{O}{6} absorbers lie within or near small (5--12 bright
galaxies) groups of \Lstar galaxies, and the remainder have only one bright galaxy
neighbor within 1~\hseventy~Mpc. The \ion{O}{6} in \lya line pairs are no
different, with only half of the sample (5/10) within or near groups of
galaxies. In both small samples, the median distance from the absorber to the
group centroid (defined by the nearest three galaxies to the absorber) is $\sim
250$~\hseventy~kpc. Of course, poorer groups may be present in the other cases, 
but with only one galaxy brighter than \Lstarno.  Better statistics may
improve these constraints.   

For the present, our best assessment is that \ion{O}{6} absorbers, particularly 
those in pairs or complexes of \lya lines, are associated with environments
of individual galaxies within a few hundred kpc. These galaxies are not 
necessarily associated with galaxy groups. Deeper galaxy survey work in 
\ion{O}{6} absorber fields is needed to search for fainter groups.

\section{Implications of Results}\label{sec:implications}

\subsection{The Extent of Metal Transport}\label{sec:transport}

The \ion{O}{6} absorbers with N$_{\rm OVI}\geq 10^{13.2}$ \cd\ are
associated with $\sim 45$\% of all \lya forest absorbers in the range
\NHI\ $=1 0^{13.2-16.5}$ \cd. Why are some of these \ion{H}{1} absorbers
detected in \ion{O}{6} and others not? No one answer may suffice. 
While our \ion{O}{6} detections are found exclusively within 800~\hseventy~kpc 
of the nearest galaxy in regions surveyed at least to \Lstarno, approximately 
two-thirds of the \ion{O}{6} non-detections are found within 1\hseventy~Mpc of 
the nearest galaxy in these same regions. It is reasonable to expect 
that \lya absorbers at greater distances from galaxies have significantly lower
metallicity, but this does not explain why many \ion{O}{6}
non-detections have nearest-neighbor distances comparable to the \ion{O}{6}
detections\footnote{We have not yet conducted a ``blind search" for \ion{O}{6} 
absorption, unassociated with \lya absorbers. In our \FUSE\ surveys, we have
not found an \ion{O}{6} absorber without some corresponding \lya absorption,   
suggesting multiphase absorbers (Danforth \& Shull 2005).}  

One likely possibility is that \ion{O}{6} absorbing gas has less than 100\% 
covering factor near galaxies. If this is the case,
then our \ion{O}{6} statistics in \S~3.2
overestimate the median distance to which metal-enriched gas is spread from
galaxies. Another possibility is that the \ion{O}{6} detections
and non-detections $\leq 800$~\hseventy~kpc from galaxies have similar
metallicities, whereas \ion{O}{6} is detected only in the most highly ionized 
systems, such as shocks with \dv $\geq 150$ \kmsno. If the \ion{O}{6} 
non-detections are systematically less ionized than the detections, 
these absorbers will exhibit associated lower ionization metal lines,
as in seen in some Galactic high-velocity clouds \citep{Sembach99,Collins04}.  
In this case, our original estimates (columns 2 and 5 of Table~\ref{nntable})
of median galaxy-absorber distances are accurate, and metals with 
$[Z]$ \simgt $-1$ have been spread to 625~\hseventy~kpc ($L \geq$ \Lstarno) 
and 335~\hseventy~kpc ($L \geq$ 0.1\Lstarno). 
In these same cases, the CDFs suggest that metals have been transported to 
maximum distances of 800 and 400~\hseventy~kpc, respectively.

These two hypotheses can be tested by searching for other metal absorption
lines (e.g., \ion{Si}{2}, \ion{S}{2}, \ion{C}{2}, \ion{Si}{3}, \ion{C}{3},
\ion{C}{4}) in \FUSE\ and \HST\ spectra. If the ionization-threshold hypothesis 
is correct, we should find lower ionization states in many absorbers where 
\ion{O}{6} is not present.  If this trend is driven primarily by metallicity, 
lower ions should be found exclusively where \ion{O}{6} has been found (DS05).  
\FUSE\ spectra have detected \ion{C}{3} $\lambda 977$ \citep{Danforthprep} 
for N$_{\rm CIII} = 10^{12.6-13.9}$ \cd. For photoionization parameters
typical of low-$z$ \lya absorbers, \FUSE\ can probe average metallicities 
$\sim 10$\% solar \citep{Danforthprep}, with sensitivities down to 
$\sim3$\% solar for strong \lya\ absorbers. 
By combining \FUSE\ detections with limits on \ion{C}{3} for photoionized gas 
and \ion{O}{6} for collisionally-ionized gas, we can assess the locations of 
10\% solar metallicity gas in the local IGM. We know of only one 
low-$z$ absorber, the 1586\kms absorber toward 3C~273 \citep{Tripp02},  
where low-ionization metal lines are detected, but \ion{O}{6} and \ion{C}{3} 
remain undetected. Thus, 
we expect that detections in either \ion{O}{6} and/or \ion{C}{3} provide a 
nearly complete metal inventory in low column density \lya absorbers. 

Table~\ref{metaltable} summarizes the statistics of metal detections in 
the DS05 sample, for \NOVI\ $\geq 10^{13.2}$ \cd\ and 
N$_{\rm CIII} \geq 10^{12.6}$ \cd, using new results from 
\citet{Danforthprep}. The nearest-neighbor statistics differ slightly in 
Table~\ref{metaltable} and Table~\ref{nntable}.  Not all \ion{O}{6} detections 
are used in Table~\ref{metaltable} if the \ion{C}{3} limits are poor, and not 
all \ion{C}{3} detections are used if the \ion{O}{6} limits are poor.
The samples in Table~\ref{metaltable} are defined more precisely, but 
the sample sizes are much smaller (see column 3).
Slightly over half of all \lya absorbers with \NHI\
$=10^{13.2-16.5}$ \cd\ have 10\% solar metallicity based upon \FUSE\
observations of \ion{C}{3} and \ion{O}{6} (DS05; Danforth \etal 2005). 
These statistics accurately assess both the frequency per unit
redshift and the proximity to galaxies of 10\% solar metallicity gas 
in the low-$z$ IGM.

For the three metal-bearing absorber samples, the median distances to 
nearest-neighbor galaxies are 100--500~\hseventy~kpc, and all three samples 
have maximum distances of $\sim 650$~\hseventy~kpc from galaxies in 
\Lstarno-complete regions.  Using either the complete \ion{C}{3} sample, 
whose nearest-neighbor statistics are shown in Table~\ref{nntable}, 
or the (9 of 16) \ion{C}{3} detections in Table~\ref{metaltable}, we find 
that no \ion{C}{3} metal line system lies farther than $\sim 800$~\hseventy~kpc 
from the nearest bright galaxy. We conclude that low column density \lya 
absorbers with nearest neighbors $> 800$~\hseventy~kpc away have $\Zmean < -1$.  
Because there are sensitive non-detections of metals within 800~\hseventy~kpc
of galaxies, regions of high metallicity cannot have a high covering factor. 
We can assess the galaxy proximity of the absorbers not detected in
metals using the 15 \ion{O}{6} absorbers in \Lstarno-complete regions from
Table~\ref{metaltable}. Fully 57\% 
of these non-detections are found $\leq 800$ \hseventy~kpc from the nearest 
known galaxy. If both the metal-rich (first three samples) and metal-poor 
(non-detections) absorbers in \Lstarno-complete regions are representative, 
then two-thirds of the absorbers with nearest-neighbor galaxies within 
800~\hseventy~kpc have $\sim 10$\% metallicity. 

We can restate this result in another way.  We can associate a volume 
extending 800~\hseventy~kpc from the nearest galaxy with the volumes 
within supercluster filaments of galaxies.  If these galaxy filaments 
were filled entirely with gas with \NHI\ $\geq 10^{13}$ \cd, then 
metal-enriched gas ($\Zmean$ \simgt\ $-1$) has a 2/3 covering factor 
in galaxy filaments.  Neither the current absorber sample nor any other 
\HST-observed sample of AGN sight lines is sufficiently dense on the sky 
to estimate the covering factor of a galaxy supercluster filament at a 
specific \ion{H}{1} column density.  Confirming the 
covering factor for \ion{H}{1} gas will have to await observations of
a much larger number of AGN targets in the far-UV. 

Nevertheless, we can still correct for \lya non-detections. As shown in 
\S~\ref{sec:nondetects}, only 50\% of all galaxies or groups 
of galaxies identified as supercluster filaments are detected at 
\NHI\ $\geq 10^{13.2}$ \cd.  
The lack of a strong correlation between lower column density 
\lya absorbers and galaxies (Figure 1) and the absence of 
clustering of lower column density \lya absorbers (Paper~IV) 
suggest that weaker absorbers are not associated with galaxies and are 
unlikely to contain metals. Thus, we suspect that both \lya 
non-detections (20\% of the total) and \lya detections at 
\NHI\ $<10^{13.2}$ \cd\ (30\% of the total) have metallicities of 
$[Z] \leq -1$. On the other hand, high-temperature 
WHIM absorbers ($T \geq 10^6$~K) are expected to be \lya non-detections but 
with observable O~VII and O~VIII \citep{Fang02,Nicastro05}. 
With these covering factors estimates, we decrease our best estimate for the 
median distance that metals are spread from \Lstar galaxies 
to (350--500)~\hseventy~kpc. Around galaxies surveyed to 0.1\Lstarno, 
the corrected distance decreases to (200--270)~\hseventy~kpc, even if these 
galaxies are not the source of the metals. The ranges indicate the uncertainty 
in covering factor arising from our lack of knowledge of metallicity in low-\NHI\ 
absorbers.

\citet{Tumfang05} used the observed frequency, $d {\cal N}/dz = 20\pm3$ (DS05), 
of \ion{O}{6} absorbers with \NOVI\ $\geq 10^{13.2}$ \cd, to constrain the 
distribution of metals in the low-$z$ IGM. From a large SDSS sample of galaxies 
and QSOs, they constructed mock catalogs of QSO/galaxy pairs.  A direct
association of galaxies and \ion{O}{6} absorbers requires a uniform (100\% 
covering factor) metal-enrichment to 750~\hseventy~kpc for $L \geq$ \Lstar
and to 300~\hseventy~kpc for $L \geq 0.1$\Lstarno.  
\citet{Danforthprep} made a similar \ion{O}{6} cross-section calculation of 
400~\hseventy~kpc ($L \geq 0.1$\Lstarno) using the observed galaxy luminosity 
function \citep{Marzke98, Blanton03}.  The difference in values 
obtained by these two approaches probably comes from galaxy clustering, 
which the \citet{Tumfang05} approach automatically takes into account.
If galaxies brighter than \Lstar (or even 0.1\Lstarno) were responsible 
for IGM metal enrichment, we would find \ion{O}{6} absorbers at greater 
distances from these galaxies.  However, a smaller number of metal-production 
sites requires larger absorber cross sections to match the observed O~VI 
line frequency, $(d{\cal N}/dz)_{\rm OVI}$.  This requirement drives us 
to propose galaxies fainter than 0.1\Lstar as the primary contributors to 
metals in the IGM \citep{Shull96}.  

For low-luminosity galaxies to produce metal-enriched absorbers, they 
must disperse these metals to 
$\sim$100~\hseventy~kpc. \citet{Stocke04} discovered one such dwarf 
galaxy-absorber pair in the 3C~273 sight line (70~\hseventy~kpc
separation on the sky), and Table 2 lists four other pairs.
In general, few absorber regions have been surveyed to sufficient depth 
to adequately test this hypothesis. Most of the entries in Table 2 are 
more luminous galaxies in regions insufficiently surveyed to
see whether dwarfs are present near these absorbers. Deeper 
searches below 0.1\Lstar are needed to uncover the other galaxies 
responsible for the bulk of the $\sim 0.1$ solar metallicity enrichment in 
the low-$z$ IGM.

\subsection{Amount of WHIM in \ion{O}{6} Absorbers}\label{sec:amount}

The difficulties of distinguishing \ion{O}{6} absorbers as collisionally 
ionized or photoionized have long been appreciated.  In many cases, 
collisional ionization is favored, such as the \ion{O}{6} absorber at 
$z = 0.1212$ toward H1821+643 \citep{Tripp01} or the \ion{O}{6}
and \ion{Ne}{8} absorber at $z = 0.2070$ toward HE0226-4110
\citep{Savage05}. In other cases \citep{Savage02, Tripp02, Prochaska04} 
an argument can be made for photoionization. The \lya profile of the 
H1821+643 absorber in Figure~\ref{OVIlineprofiles} shows absorption spread 
over 75 \kmsno. \citet{Tripp01} suggest that this radial velocity 
implies a full shock velocity $\sim 130$ \kmsno, within the constraints 
required to produce the observed \ion{O}{6} and \ion{H}{1} line strengths 
and widths, as well as the upper limit on \ion{C}{4} in the case of
collisional ionization.   

In collisional ionization equilibrium, \ion{O}{6} reaches its maximum 
ionization fraction at $T = 10^{5.45}$~K \citep{SuthDop93}. In the
non-equilibrium cooling regions behind radiative shocks, \ion{O}{6}
forms in detectable quantities for shock velocities $V_s \geq 130$ \kms
\citep{SM79,IndShu04}, with substantial amounts produced for $V_s = 150-200$ 
\kms.  The post-shock temperature at an adiabatic shock front is,
\begin{equation}
   T_s = \left( \frac {3 \mu V_s^2}{16 k_B} \right) \approx
        (1.34 \times 10^5~{\rm K}) \left( \frac {V_s}{100~\kmsno} 
         \right)^2  \; ,      
\end{equation}
where $\mu \approx 0.59 m_H$ is the mean particle mass for a fully
ionized plasma with He/H = 0.08.  Generalizing this result to all \ion{O}{6} 
absorbers, we expect shock-heated \ion{O}{6} absorbers to be members of 
\lya line pairs with \dv $= 75-150$ \kmsno, adopting mean shock velocities 
$\sqrt{3}$ times the radial velocity differences. 
We found that a substantial fraction of \ion{O}{6} absorbers correspond to
paired or complex \lya absorption lines with separations \dv $\leq 200$ \kmsno.
These \ion{O}{6} absorbers in \lya line pairs have nearest \Lstar galaxies
ranging from 140--450~\hseventy~kpc, significantly less than for the full
\ion{O}{6} sample. 

This distinction of paired or non-paired absorbers may be important, since  
\lya line pairs with one or more detections of \ion{O}{6} are 
potential shock-heated absorbers. Approximately 45\% of the
\ion{O}{6} detections occur in \lya line pairs, and \ion{O}{6} absorbers  
are seen in 45\% of the higher column density \lya absorbers.  In \S~3.4
we suggested that this subset (20\% of the stronger \lya absorbers)
provides the best candidates for shock-heated WHIM.  They typically 
show \ion{O}{6} absorption in one line of paired \lya absorbers.  
A somewhat larger number is obtained using the \ion{O}{6} and \ion{C}{3} 
statistics in Table~\ref{metaltable}, which shows that
half the metal-line detections appear in both \ion{O}{6} and \ion{C}{3}
\citep{Danforthprep}.  Other shock-heated systems may have 
smaller radial velocity components along the sightline.
These estimates are made more difficult 
by the fact that many absorbers contain multi-phase components of 
collisionally-ionized and photoionized gas. 

The number of \ion{O}{6} absorbers in
\lya line pairs is sufficient to explain the excess power in the TPCF of the
low-$z$ \lya forest at \dv $\leq 200$\kms (Paper~IV). The \ion{O}{6} absorbers
in \lya line pairs also have the nearest galaxies of all the samples compiled
in this paper. These distances are significantly less than for the
\ion{O}{6} absorber sample as a whole, and they are as expected for WHIM on the
basis of the numerical simulations \citep{Dave99}. The \lya line doubling
suggests models for these WHIM absorbers involving either primordial infall
onto metal-enriched galaxy halos or metal-enriched galactic winds impacting 
primordial IGM clouds.  Because these systems involve more than one absorber, 
mixed-ionization (multiphase) models that form both \ion{C}{3} and \ion{O}{6} 
may naturally result. The mixed ionization and close proximity to galaxies 
suggest a comparison between these 
absorbers and the Galactic HVCs \citep{Collins05, Fox05}.

DS05 used the current sample of \ion{O}{6} absorbers to determine the 
WHIM contribution of the \ion{O}{6} absorbers to the baryon density,  
\begin{equation}
   \Omega_b({\rm OVI}) = (0.0022 \pm 0.0003) \left[ h_{70} (Z_O/Z_{\odot}) 
     (f_{\rm OVI}/0.2) \right]^{-1} \; ,
\end{equation}
This represents a baryon fraction, $\Omega_b(\rm WHIM)/ \Omega_b(\rm tot) 
\approx 5$\%, assuming that this gas has 10\% solar metallicity and 
ionization fraction $f_{\rm OVI} = 0.2$. 
The \ion{O}{6} contribution to $\Omega_b$ could be as high as 10\%, 
considering the likely dispersion in O/H metallicity and \ion{O}{6} ionization 
fraction. Using an \HST\ sample, \citet{Tripp05} found a 
similar fraction of baryons (7\%) in \ion{O}{6} absorbers at $z \geq 0.15$. 

Some of the \ion{O}{6} absorbers may include a 
contribution from photoionization, which raises the issue of
``double counting" in the baryon inventory (Stocke, Shull, \& Penton 2005). 
In our \HST\ surveys of \lya absorbers (Papers I and IV), we converted from
observed H~I column densities to the total number of baryons by
assuming photoionization equilibrium in warm ($\sim10^4$~K) mostly
ionized gas.  In contrast, the \ion{O}{6} surveys assume collisionally
ionized gas at $\sim 10^{5.5}$~K.  Because the H~I fractions in this hot 
gas are low, $\sim 10^{-6}$, any \lya absorption will be broad 
(FWHM $\approx 120$ \kms) and shallow (optical depth $\tau_0 \approx 0.1$),
in contrast to the observed \lya lines, which are typically sharp and narrow
($b < 40$ \kmsno).  For these reasons, we believe that the \lya surveys 
{\it do not} count H~I absorption from the WHIM.   One might expect a 
small reduction in the \ion{O}{6} WHIM baryon census, owing to photoionized 
\ion{O}{6} absorbers that co-exist with single-phase, photoionized \lya lines.  

It is also worth elaborating on how the \lya and \ion{O}{6} surveys were 
done.  Both DS05 and \citet{Danforthprep} 
used the surveyed \lya lines as ``signposts" in redshift to search 
for corresponding \ion{O}{6} and \ion{C}{3}. The fact that they found
\ion{O}{6} in a significant fraction of cases shows that these ions 
are associated kinematically with the H~I.  Physical modeling suggests
that these ions exist in a multi-phase medium, in which the hot gas and 
warm photoionized gas share the same velocity range, but are not 
cospatial.  Such situations could arise in conductive interfaces, 
in cooling shocks, or in turbulent mixing layers.  In our baryon 
surveys, the \lya census counts warm photoionized baryons (30\% of
$\Omega_b$) and the \ion{O}{6} census counts a portion of the hot 
(WHIM) baryons (5--10\% of $\Omega_b$).

Our arguments in \S~4.2  about shock-heating suggest that \ion{O}{6} absorbers 
in \lya pairs are collisionally-ionized gas at $T = 10^{5-6}$~K.  The other 
half of the \ion{O}{6} sample probably also includes shock-heated gas 
with low radial-velocity components, as well as some low-density 
($n_H\sim 10^{-5.5}$ cm$^{-3}$) photoionized ($T \sim 10^4$~K) gas.
These absorbers would be similar to, but much more extended than typical 
\lya absorbers \citep[see][for sample calculations]{Tripp01, Danforthprep}. 
With much better statistics, one could quantify the appropriate reduction in 
\ion{O}{6}-associated baryon counts, owing to those \ion{O}{6} 
absorbers produced in single-phase (warm) photoionized gas.

\section{Conclusions and Discussion}\label{sec:conclusions}

We have used the \ion{O}{6} sample of DS05, with 37 detections at \NOVI\ $\geq
10^{13.2}$ \cd\ and 45 non-detections with upper limits below that level,  
to investigate the possible association of \ion{O}{6} absorbers with galaxies. 
Our sample of \ion{O}{6} absorbers comes from the stronger half of 
\lya absorbers (Papers~I and IV).
The redshifts of these absorbers ($z < 0.15$) allow us to compare
their locations with 1.07 million galaxy redshifts and positions from 
the CfA (Jan. 2005) catalog, the Sloan Digital Sky Survey (DR4), and 
the 2dF-FDR, 6dF-DR2, and Veron-11 catalogs. 
In regions where the galaxy surveys are complete to at least \Lstarno,
we found 23 \ion{O}{6} detections and 32 non-detections.

\noindent
The major results of our study can be summarized as follows:
\begin{enumerate}

\item We confirm that stronger \lya absorbers (\NHI\ $=10^{13.2-16.5}$ \cd) 
   are three times closer to ($\geq0.1$ \Lstarno) galaxies, on average, than 
   weaker absorbers.  Nearest-neighbor statistics suggest that these absorbers 
   are related to galaxy supercluster filaments. 

\item Among the stronger \lya lines, we found 10 \lya absorbers in voids
  ($\sim15$\% of the total), with nearest-neighbor galaxies 
  more than 3~\hseventy~Mpc away.  Interestingly, none of these void 
  absorbers were detected in \ion{O}{6}.

\item The \ion{O}{6} and \ion{C}{3} absorbers nearly always lie within 
  800~\hseventy~kpc of galaxies in \Lstarno-complete regions.  Nearest-neighbor 
  statistics place these absorbers sufficiently close to 
  individual galaxies that they could be associated with winds or halos.
  Correcting for covering factors, we estimate that metals have been
  spread to median distances (350--500)~\hseventy~kpc from \Lstar galaxies
  and (200--270)~\hseventy~kpc from 0.1\Lstar galaxies.

\item A substantial percentage (16/37= 43\%) of \ion{O}{6} absorbers are 
   associated with \lya absorption in close pairs with \dv $= 50-200$ \kmsno.  
   This velocity difference suggests a physical origin of some \ion{O}{6} 
   absorbers in shock-heated gas.   

\end{enumerate}

We now elaborate on these overall conclusions and their implications.  
In general, the median distance between any two $\geq 0.1$\Lstar galaxies 
is less than the median distance from the absorbers to the nearest 
$\geq 0.1$\Lstar galaxy (Table~\ref{nntable}, column~5). Thus, it is 
often difficult to determine which galaxy, if any, is the source 
of the gas in the absorber, a result consistent with numerical simulations 
\citep{Dave99}.  A more sensible picture places the \lya absorbers within
supercluster filaments.  The nearest-neighbor statistics for both
\ion{O}{6} and \ion{C}{3} absorbers place them sufficiently close to 
individual galaxies that some could be associated with winds or halos.
Because few absorber regions have been surveyed below 0.1\Lstarno, we cannot 
exclude the possibility that metal-line absorbers are associated with 
undiscovered faint galaxies.  Some very faint galaxies have been found 
close to individual metal-line absorbers \citep[Table~\ref{assoctable} 
;][]{vanGorkom96, Shull96, Stocke05} suggesting that dwarf
galaxy superwinds could be responsible for many weak metal-line systems.
New galaxy survey work around the very nearest ($cz\leq 5000$ \kmsno) absorbers
is underway to test this possibility. 

The \ion{O}{6} and \ion{C}{3} detections correspond to \simgt 10\% of solar 
metallicity for plausible models of collisional ionization and photoionization, 
respectively.  The statistics of this sample suggest that metals have 
been spread to median distances of 350--500~\hseventy~kpc from \Lstar galaxies 
and 200--270~\hseventy~kpc of 0.1\Lstar galaxies, regardless of whether these
galaxies are the source of these metals. The quoted range reflects
the presence of regions of uncertain metallicity when the \lya absorption is
weak or undetected.
The extent of metal transport must be variable, since \ion{C}{3}
and \ion{O}{6} detections and non-detections are found at comparable distances
from galaxies. Weak \lya (\NHI\ $< 10^{13.2}$ \cd) detections and
non-detections are present in our sample with similar nearest-neighbor
distances, but we do not know their metallicity.

The \ion{O}{6} absorbers associated with close pairs of \lya with 
\dv $= 50-200$\kms imply velocities in the correct range to produce large 
amounts of \ion{O}{6} in shocks \citep{Shull03, Tumlinson05}. 
About half of all \ion{O}{6} absorbers, and half the \ion{O}{6}
absorbers in \lya line pairs, are found in or near groups of galaxies.
In the \lya absorber complex toward PKS~2155-304 \citep{Shull98, Shull03},
some of the absorbers contain \ion{O}{6} in apparent association with
galaxy groups.  Dwarf galaxies with 
$\sim 100$~\hseventy~kpc halos or winds will have that gas stripped by 
frequent collisions with other dwarfs in $\simlt 10^9$ yr.  For the 
\ion{O}{6} absorbers in or near galaxy groups, a nearby dwarf may not be 
present, having been stripped of its gas to make a more general intragroup 
medium, as envisioned originally by \citet{Mulchaey96}.

The \ion{O}{6}/\lya pairs are found closer to galaxies, by about a factor of 
two compared to the overall \ion{O}{6} absorber sample from which they were 
chosen.  The median distance between any one of these absorbers and a galaxy 
of {\it any} luminosity in \Lstarno-complete regions is 190~\hseventy~kpc, 
from which we extrapolate that most of these \ion{O}{6} absorbers are $\leq
100$~\hseventy~kpc from galaxies at 0.1\Lstar or fainter. These 
nearest-galaxy distances are a close match to galaxy distances from
collisionally-ionized WHIM in the \citet{Dave99} simulations. Given 
their relationship to other nearby absorbers and to nearby galaxies, we
interpret the \ion{O}{6}/\lya line pairs as shocked, collisionally-ionized gas.
This type of \ion{O}{6} absorber shares some observational similarities with
Galactic HVCs \citep{Sembach03,Collins04, Fox05}. The \ion{O}{6} sample in 
\lya pairs is substantial, with 5 pairs per unit redshift at low-$z$.
These \ion{O}{6} absorptions are sufficient to account for most or all of 
the excess power seen in the two-point correlation function of
\lya absorbers (Paper~IV).  

Guided by the maximum extent to which metals are spread from galaxies, we 
define a filament-absorber association for systems $\leq 800$~\hseventy~kpc 
from the nearest galaxy, in regions surveyed down to at least \Lstarno. 
Because filaments are typically 5--7~\hseventy~Mpc across \citep{Ramella92}, 
the nearest galaxy usually lies well within the filament.  Our estimate of 
800~\hseventy~kpc is a reasonable choice, given the average density of 
galaxies within a filament.  From our results of \S~3.3, we envision an 
empirical model for a supercluster filament in the local universe,  
composed of three generic regions: 
(1) Region \#1 (33--50\% coverage) composed of metal-enriched 
($[Z]$ \simgt $-1$) absorbers; 
(2) Regions \#2 (16--25\% coverage) of low metallicity absorbers; and 
(3) Regions \#3 (25--50\% coverage) of unknown metallicity, owing to weak 
or non-detections of \lyano.  

Our previous work led us to conclude that weaker \lya absorbers are only 
loosely associated with galaxies (see Figure 1) and may represent a population 
of uniformly distributed absorbers (Paper~III). We propose here that the weaker 
\lya absorbers lie in regions of low-metallicity and should be added to the 
accounting of Region \#2.
In $10^6$~K gas, with neutral fraction $f_{\rm HI} \approx 10^{-6}$,  
the WHIM produces broad, shallow \lya absorption.  Therefore, the \lya 
non-detections might be metal-rich regions too hot to be seen in \lya
absorption, but visible in {\it Chandra} spectra of high-ionization 
metal-line absorptions \citep{Fang02, Nicastro05}. 
The extent of metal transport calculated above refers to the spread of
gas of $0.1 Z_{\odot}$ metallicity from galaxies in the current epoch. 
Absorbers more than 1~\hseventy~Mpc from galaxies may all contain gas at
$\Zmean \approx -1.5$ in the current epoch or they may be metal-free.
We do not address the presence or absence of very metal-poor gas \citep[e.g.,
absorbers of 3\% solar metallicity at high-$z$;][]{Aguirre02}.

\acknowledgements
The authors acknowledge several \HST\ and \FUSE\ grants to the University of
Colorado in support of this work: HST-GO-06593.01, GO-08182.01, AR-09221.01,
FUSE NNG04GA18G, the HST/COS project (NAS5-98043), NASA NAG5-7262, NSF
AST02-06042, FUSE NAG5-13004. J.T. acknowledges partial support from the
Department of Astronomy \& Astrophysics at the University of Chicago and 
HST-G0-9874.01-A. We thank John Huchra and Nathalie Martimbeau for providing 
timely updates to the CfA Redshift Survey Catalog, R.~J. Weymann for 
his assistance with galaxy survey work, and Mark Giroux for comments
on the manuscript.
This work was based partially on observations obtained with the Apache 
Point Observatory 3.5-meter telescope, which is owned and operated by the 
Astrophysical Research Consortium.
Funding for the creation and distribution of the SDSS Archive 
(\url{http://www.sdss.org/}) has been provided
by the Alfred P. Sloan Foundation, the Participating Institutions, NASA, NSF,
the U.S. Department of Energy, the Japanese Monbukagakusho, and the Max Planck
Society. 

\newpage

\bibliographystyle{apj}


\begin{deluxetable}{l|rrr|rrr}
\tableheadfrac{0.01}
\tablecolumns{7}
\tablecaption{Median Nearest-Neighbor Distances\label{nntable}}
\tablewidth{0pt}
\tabletypesize{
\scriptsize}
\tablehead{
\multicolumn{1}{c}{} &
\multicolumn{3}{c|}{\Lstarno-complete Regions} & \multicolumn{3}{c}
    {0.1\Lstarno-complete Regions} \\
\multicolumn{1}{c|}{Sample} &
\multicolumn{1}{c}{$\geq $\Lstarno} &
\multicolumn{1}{c}{All Galaxies\tablenotemark{a}} &
\multicolumn{1}{c|}{N\tablenotemark{b}} &
\multicolumn{1}{c}{$\geq 0.1$\Lstarno} &
\multicolumn{1}{c}{All Galaxies\tablenotemark{c}} &
\multicolumn{1}{c}{N\tablenotemark{b}} \\
\multicolumn{1}{c|}{} &
\multicolumn{1}{c}{(\hseventy~kpc)} &
\multicolumn{1}{c}{(\hseventy~kpc)} &
\multicolumn{1}{c}{} &
\multicolumn{1}{|c}{(\hseventy~kpc)} &
\multicolumn{1}{c}{(\hseventy~kpc)} &
\multicolumn{1}{c}{}
}
\startdata
Galaxy - Galaxy& 1780 & 345 & 480 & 245 &120&630\\
\lya - Stronger Half\tablenotemark{d} & 1130 & 450 & 69 & 445 & 280 &20\\
\lya - Weaker Half & 2150 & 1855 & 69 & 1455 & 480 &19\\
O~VI Absorbers\tablenotemark{e} & 625 & 285 & 23 & 335 &180&9\\
O~VI Upper Limit & 1090 & 500 & 32 & 385 & 320 &8\\
O~VI Absorbers in \lya Pairs\tablenotemark{e} & 290 & 190 & 10 &
\nodata\tablenotemark{g} & \nodata\tablenotemark{g} & 3 \\ 
C~III Absorbers & 430 & 150& 13 & \nodata\tablenotemark{g} &
\nodata\tablenotemark{g} & 3\\ 
\enddata
\tablenotetext{a}{Using all available galaxies, of any luminosity,
     in regions complete to at least \Lstarno.}
\tablenotetext{b}{Number of galaxies or absorbers in sample.}
\tablenotetext{c}{Using all available galaxies, of any lumonosity,
    in regions complete to at least 0.1\Lstarno.}
\tablenotetext{d}{W$_{\lambda} \geq 68$~m\AA\ or log~\NHI\ $\geq 13.1$ for \lya.}
\tablenotetext{e}{\FUSE\ O~VI absorber sample from \citet{Danforth05}.}
\tablenotetext{f}{This sample is described in the text below.}
\tablenotetext{g}{Too few absorbers in 0.1\Lstar regions to report accurate median.}
\end{deluxetable}


\begin{deluxetable}{lrrlrrrrr}
\tableheadfrac{0.01}
\tablecolumns{9}
\tablecaption{Galaxies within 200~\hseventy~kpc of \lya/O~VI detections  
\label{assoctable}}
\tablewidth{0pt}
\tabletypesize{\scriptsize}
\tablehead{
\colhead{Target} &
\colhead{V$_{\rm abs}$\tablenotemark{a}} &
\colhead{\ensuremath{W}(\lya)\tablenotemark{b}} &
\colhead{Galaxy\tablenotemark{c}} &
\colhead{$m_B$\tablenotemark{d}} &
\colhead{$M_B$\tablenotemark{e}} &
\colhead{V$_{\rm gal}$\tablenotemark{h}} &
\colhead{Galaxy RA,DEC} &
\colhead{D$_{\rm tot}$\tablenotemark{f}} \\
\colhead{} &
\colhead{({\rm km~s}\ensuremath{^{-1}})} &
\colhead{(m\AA)} &
\colhead{Name} &
\colhead{}&
\colhead{}&
\colhead{({\rm km~s}\ensuremath{^{-1}})} &
\colhead{(J2000)} &
\colhead{}
}
\startdata
3C~273& 1015& 369&SDSS-O587726031714779292& 16.3&-14.3& 903& 12 28
16.0 +01 49 43.9& 69\\
3C~273& 1015& 369&12285+0157 & 15.6&-15.4& 1105& 12 31 03.8 +01 40
34.2& 169\\
H~1821+643& 36394& 390&A1821+6420H & 18.3&-20.3& 36436& 18 22 02.7
+64 21 38.7& 162\\
Mrk~335&1966& 220& 00025+1956 &  16.0& -16.2&  1950&       00 05
29.4  +20 13 36.0&   95\\
Mrk~876& 958& 324&N6140 & 12.6&-18.0& 910& 16 20 56.9 +65 23 21.7& 180\\
Mrk~876& 958& 324&16224+6533 & 16.5&-13.9& 822& 16 22 49.7 +65 26
07.2& 196\\
PG~1116+215& 41579& 476&A1116+2134A & 18.4&-20.5& 41467& 11 19 06.8
+21 18 28.8& 139\\
PG~1211+143& 2130& 186&I3061 & 14.9&-17.7& 2317& 12 15 04.7 +14 01
40.7& 110\\
PG~1211+143& 15384& 999&McLin~396 & 16.7&-20.0& 15242& 12 14 09.6 +14
04 21.1& 136\\
PG~1259+593& 13818& 685&SDSS-O587732590107885591& 16.9&-19.6& 13856&
13 01 01.0 +59 00 06.7& 138\\
PG~1259+593& 13950& 463&SDSS-O587732590107885591& 16.9&-19.6& 13856&
13 01 01.0 +59 00 06.7& 138\\
Ton~S180& 7017& 222&McLin~564 & 19.9&-15.1& 6980& 00 57 04.2 -22 26
58.2& 153\\
\enddata
\tablenotetext{a}{Heliocentric velocity ($cz$) of detected O~VI 
   and \lya absorber (\NHI\ $\geq 10^{13.2}$ \cd).  .}
\tablenotetext{b}{Equivalent width of detected (4$\sigma$) \lya
    absorber}
\tablenotetext{c}{Name of non-detected galaxy. McLin\# indicates that
   the galaxy is taken from \citet{McLin02}.}
\tablenotetext{d}{Apparent blue (Zwicky) magnitude of galaxy.}
\tablenotetext{e}{Absolute blue (Zwicky) magnitude of galaxy. Our
   definition of an \Lstar galaxy is $M_B =-19.57$.}
\tablenotetext{f}{Total Euclidean distance (in \hseventy~kpc) to the
  nearest absorber assuming our ``500 \kms retarded Hubble flow'' model.}
\tablenotetext{h}{Heliocentric velocity ($cz$) of the galaxy.}
\end{deluxetable}


\begin{deluxetable}{rcccrrrrr}
\tableheadfrac{0.01}
\tablecolumns{9}
\tablecaption{Non and Weak \lya Detections near Galaxies within 200~
\hseventy~kpc of Sight Lines.}
\label{nontable}
\tablewidth{0pt}
\tabletypesize{\scriptsize}
\tablehead{
\colhead{Target} &
\colhead{V$_{abs}$\tablenotemark{a}} &
\colhead{\ensuremath{W}\tablenotemark{b}} &
\colhead{Galaxy\tablenotemark{c}} &
\colhead{$m_B$\tablenotemark{d}} &
\colhead{$M_B$\tablenotemark{e}} &
\colhead{V$_{gal}$\tablenotemark{h}} &
\colhead{Galaxy RA,DEC} &
\colhead{D$_{tot}$\tablenotemark{f}} \\
\colhead{} &
\colhead{({\rm km~s}\ensuremath{^{-1}})} &
\colhead{(m\AA)} &
\colhead{Name} &
\colhead{}&
\colhead{}&
\colhead{({\rm km~s}\ensuremath{^{-1}})} &
\colhead{(J2000)} &
\colhead{}
}
\startdata
3C~273 & \nodata& $< 38$ & McLin~3 & 19.4& -14.0& 3328 &12 28 21.6
+01 56 46.8& 176\\
Mrk~817  &1933&      29& SDSS-O588011219675513024&  16.9& -15.1&
1768&       14 39 03.9  +58 47 17.2&  153\\
Mrk~1383 & 8951& 66& McLin~243                     &  19.5& -16.0&
9002&       14 28 58.7  +01 13 08.0&  160\\
PG~0804+761 & 1530&      78& 08050+7635  &  13.5& -18.2&  1544&  08
11 37.0  +76 25 16.6&  144\\
PG~1211+143 & \nodata& $<45$& I3065 & 14.7& -16.2& 1072 &12 15 12.3
+14 25 57.7& 116\\
PG~1211+143 & \nodata& $<43$& SDSS-O588017567636652218     &  16.3&
-15.1&  1305& 12 14 14.4  +13 32 34.4&  165\\
PG~1211+143 & \nodata& $<41$& I3077 & 15.4& -16.1& 1411 &12 15 56.4
+14 25 59.9& 192\\
PKS~2005-489&2753&      24& N6861E                       &  14.5&
-18.3&  2531&       20 11 01.2  -48 41 27.3&  186\\
\enddata
\tablenotetext{a}{Heliocentric velocity ($cz$) of detected absorber; 
   blank entry is for \lya non-detections.}
\tablenotetext{b}{Equivalent width of detected (4$\sigma$) \lya
   absorber or 3$\sigma$ limit on detection}
\tablenotetext{c}{Name of non-detected or weakly-detected galaxy.
    McLin\# indicates that the galaxy is taken from \citet{McLin02}.}
\tablenotetext{d}{Apparent blue (Zwicky) magnitude of galaxy.}
\tablenotetext{e}{Absolute blue (Zwicky) magnitude of galaxy.}
\tablenotetext{f}{Total Euclidean distance (in \hseventy~kpc) to the
   nearest absorber, assuming our ``500 \kms retarded Hubble flow'' model 
   or, for non-detections, the distance to the sight line.}
\tablenotetext{h}{Heliocentric velocity ($cz$) of the galaxy.}
\end{deluxetable}


\begin{deluxetable}{rcrc}
\tablecolumns{4}
\tablecaption{Metal-Line Absorbers \label{metaltable}}
\tablewidth{0pt}
\tabletypesize{\scriptsize}
\tablehead{
\colhead{Sample\tablenotemark{a}} &
\colhead{Total Number} & \colhead{Number in} & \colhead{Median NND\tablenotemark{b}} \\
\colhead{Description} & \colhead{in Sample} & \colhead{\Lstar Regions}
&\colhead{(\hseventy~kpc)} }
\startdata
O~VI and C~III Detections & 12 & 6 & 145 \\
O~VI Detection/C~III Upper Limit & 8 & 5 & 540 \\
C~III Detection/O~VI Upper Limit & 4 & 3 & 100\tablenotemark{c}  \\
O~VI and C~III Non-Detections & 21 & 15 & 1000 \\
\enddata
\tablenotetext{a}{Detections and non-detections defined using column density 
  limits N$_{\rm CIII}\geq 10^{12.6}$ \cd\ and N$_{\rm OVI}\geq 10^{13.2}$ \cd. }
\tablenotetext{b}{Nearest-neighbor distance (NND) using ``all galaxies" 
   method, as surveyed in \Lstarno-complete regions}
\tablenotetext{c}{Note large uncertainty owing to small-number statistics. } 

\end{deluxetable}

\end{document}